\documentclass[sigconf,10pt,nonacm]{acmart}
\usepackage{tikz}
\usepackage{amsmath}

\usepackage{amssymb}
\usepackage{listings}
\usepackage{xcolor}
\usepackage{enumitem}
\usepackage{booktabs}
\usepackage{pgfplots}
\usepackage{subcaption}
\usepgfplotslibrary{fillbetween}

\newif\ifcomments
\commentsfalse 

\ifcomments
  \newcommand{\ioannis}[1]{\textcolor{orange}{\textsf{(Ioannis: #1)}}}
  \newcommand{\tibor}[1]{\textcolor{blue}{\textsf{(Tibor: #1)}}}
  \newcommand{\laurent}[1]{\textcolor{magenta}{\textsf{(Laurent: #1)}}}
\else
  \newcommand{\ioannis}[1]{}
  \newcommand{\tibor}[1]{}
  \newcommand{\laurent}[1]{}
\fi

\definecolor{codebg}{HTML}{F7F8FA}
\definecolor{codekw}{HTML}{0B6FA4}
\definecolor{codecmt}{HTML}{6A737D}
\definecolor{codestr}{HTML}{067D17}
\definecolor{codebg}{HTML}{F7F8FA}
\definecolor{nexus}{HTML}{175E4C}
\definecolor{gapred}{HTML}{C43C3C}
\definecolor{convgreen}{HTML}{0A7A0A}
\definecolor{gridgray}{HTML}{DBDBD5}
\definecolor{inkorange}{RGB}{214,105,26}  
\definecolor{evolvegreen}{RGB}{84,124,93} 

\pgfplotsset{
  panelaxis/.style={
    width=\linewidth, height=3.0cm,
    scale only axis=false,
    axis lines=left,
    axis line style={-, line width=0.5pt, black!70},
    tick align=outside, tick style={black!70},
    x tick label style={font=\scriptsize},
    y tick label style={font=\scriptsize},
    ymajorgrids, grid style={gridgray, line width=0.3pt},
    major tick length=1.8pt,
    label style={font=\scriptsize},
    xlabel style={yshift=2pt},
    ylabel style={yshift=-4pt},
    clip=false,
    every axis plot/.append style={line width=0.8pt},
  },
}
\tikzset{
  sol/.style={nexus, mark=*, mark size=1.15pt, solid},
  prop/.style={nexus, densely dashed, line width=0.6pt, mark=o, mark size=1.15pt},
  fail/.style={gapred, font=\scriptsize, inner sep=1pt},
  conv/.style={only marks, mark=*, mark size=2.0pt, convgreen},
  convlab/.style={convgreen, font=\scriptsize\bfseries, inner sep=1pt},
}

\lstdefinelanguage{Rust}{
  morekeywords={let,mut,fn,for,in,if,else,return,self},
  morekeywords=[2]{not,and,or,implies,assume,clone,push,le,lt,equals,is_some,is_none},
  keywordstyle=\color{codekw}\bfseries,
  keywordstyle=[2]\color{codekw},
  comment=[l]{//}, morecomment=[s]{/*}{*/},
  commentstyle=\color{codecmt}\itshape,
  morestring=[b]", stringstyle=\color{codestr},
}

\lstdefinestyle{snippet}{
  language=Rust,
  basicstyle=\ttfamily\footnotesize,
  backgroundcolor=\color{codebg},
  frame=single, framerule=0pt, rulecolor=\color{codebg}, framesep=5pt,
  xleftmargin=3pt, xrightmargin=3pt,
  aboveskip=4pt, belowskip=2pt,
  columns=fullflexible, keepspaces=true,
  breaklines=true, breakatwhitespace=true,
  showstringspaces=false, numbers=none,
}

\usetikzlibrary{positioning, arrows.meta, calc, fit, backgrounds}
\definecolor{panelbg}{RGB}{237,242,250}
\definecolor{paneledge}{RGB}{201,214,234}
\definecolor{boxbg}{RGB}{244,245,247}
\definecolor{boxedge}{RGB}{203,207,216}
\definecolor{inkblue}{RGB}{43,101,179}
\definecolor{duskpurple}{RGB}{159,79,140}
\definecolor{titlegray}{RGB}{92,92,112}
\definecolor{gridgray}{RGB}{150,150,160}
\definecolor{accorange}{RGB}{228,116,58}

\newcommand{\fakepar}[1]{\vspace{0.4em}\noindent\textbf{#1}}

\title{Self-evolving network verifiers}

\author{Ioannis Protogeros}
\affiliation{%
  \institution{ETH Z\"{u}rich}%
  \city{Z\"{u}rich}
  \country{Switzerland}%
}
\email{iprotogeros@ethz.ch}
\author{Tibor Schneider}
\affiliation{%
  \institution{ETH Z\"{u}rich}%
  \city{Z\"{u}rich}
  \country{Switzerland}%
}
\email{sctibor@ethz.ch}
\author{Laurent Vanbever}
\affiliation{%
  \institution{ETH Z\"{u}rich}%
  \city{Z\"{u}rich}
  \country{Switzerland}%
}
\email{lvanbever@ethz.ch}

\begin{abstract}

Symbolic network verifiers can reason about correctness across vast spaces of routing inputs and failures, but only for the protocols and features an expert has encoded by hand. Creating and maintaining a faithful model of the control plane is both difficult and never-ending, since no written source specifies perfectly what a network does: vendor implementations deviate from the RFCs, and behaviour shifts with releases. The burden of constant upkeep ultimately keeps verification out of many networks that need it.

We argue that the model should instead evolve automatically to faithfully capture the actual network behaviour. To achieve that, we leverage the only source that specifies it unambiguously: the router software itself. In a counterexample-guided loop, a coding agent proposes extensions to the verifier's symbolic encoding, while a trusted oracle (e.g., emulated routers) supplies the ground-truth routing state. The agent iteratively refines the network model using each disagreement with the oracle.

As early evidence, a prototype of this system taught a 3,000-line SMT-based verifier three features it did not support: OSPF areas, BGP route reflection, and L3VPN over EVPN, converging autonomously on models that match the oracle, even noticing vendor-specific behaviour. Automating model growth shifts the hard problem from writing verification systems to systematically testing them; we propose a research agenda for trusting and harnessing automatically evolved verifiers.

\end{abstract}

\begin{document}

\maketitle

\section{Introduction}
\label{sec:intro}

\begin{figure}[t]
  \centering
  \scalebox{0.99}{\begin{tikzpicture}[
    font=\scriptsize,
    space/.style={draw=black!40, line width=0.5pt},
    core/.style={fill=inkblue!80, draw=none},
    grownfill/.style={fill=inkblue!18, draw=none},
    frontier/.style={draw=inkblue!70, dashed, line width=0.7pt},
    sample/.style={text=inkorange, font=\footnotesize, inner sep=0pt},
    expand/.style={-{Stealth[length=1.5mm]}, inkblue!70, line width=0.7pt},
    axis/.style={-{Stealth[length=1.2mm]}, black!55, line width=0.5pt},
    axlbl/.style={font=\scriptsize, text=black!60, inner sep=0.5pt},
    title/.style={font=\scriptsize\bfseries, inner sep=1pt, anchor=base},
    verdict/.style={font=\scriptsize\itshape, text=titlegray, align=center, inner sep=1pt, anchor=north},
  ]
  \draw[axis] (-0.42,0.30) -- (-0.42,1.15);
  \node[axlbl] at (-0.42,1.32) {$\mathcal{E}$};
  \draw[axis] (-0.42,0.30) -- (-0.04,0.30);
  \node[axlbl] at (-0.20,0.14) {$\mathcal{C}$};
  \node[axlbl, anchor=base] at (3.925,2.02)
    {$\mathcal{C}$: configuration space \quad\;\; $\mathcal{E}$: environment space};
  \begin{scope}
    \fill[core] (0.80,0) rectangle (1.50,1.9);
    \draw[space] (0,0) rectangle (2.3,1.9);
    \node[title]   at (1.15,-0.36) {control-plane verification};
    \node[verdict] at (1.15,-0.60) {all environments within\\ a fixed configuration space};
  \end{scope}
  \begin{scope}[xshift=2.75cm]
    \draw[space] (0,0) rectangle (2.3,1.9);
    \foreach \p in {(0.35,0.40),(0.60,1.45),(1.05,0.90),(1.55,1.55),
                    (1.90,0.35),(2.05,1.10),(0.95,0.25),(1.40,1.15),
                    (0.30,1.00),(1.75,0.75),(0.80,1.70)}
      \node[sample] at \p {$\ast$};
    \node[title]   at (1.15,-0.36) {emulation};
    \node[verdict] at (1.15,-0.60) {any configuration,\\ one environment at a time};
  \end{scope}
  \begin{scope}[xshift=5.5cm]
    \fill[grownfill] (0.32,0) rectangle (1.98,1.9);
    \fill[core] (0.80,0) rectangle (1.50,1.9);
    \draw[frontier] (0.32,0) -- (0.32,1.9);
    \draw[frontier] (1.98,0) -- (1.98,1.9);
    \draw[space] (0,0) rectangle (2.3,1.9);
    \draw[expand] (1.52,1.35) -- (1.90,1.35);
    \draw[expand] (1.52,0.55) -- (1.90,0.55);
    \draw[expand] (0.78,1.35) -- (0.40,1.35);
    \draw[expand] (0.78,0.55) -- (0.40,0.55);
    \foreach \p in {(0.52,1.62),(0.58,0.85),(0.50,0.22),
                    (1.68,1.68),(1.62,0.72),(1.74,0.28)}
      \node[sample] at \p {$\ast$};
    \node[sample] at (1.98,0.95) {$\ast$};
    \node[title]   at (1.15,-0.36) {self-evolving};
    \node[verdict] at (1.15,-0.60) {space grown through\\ oracle-checked scenarios};
  \end{scope}
\end{tikzpicture}}
  \caption{A self-evolving verifier keeps the symbolic model's guarantees
  and grows their coverage, guided by ground-truth observations.}
  \label{fig:tradeoff}
\end{figure}
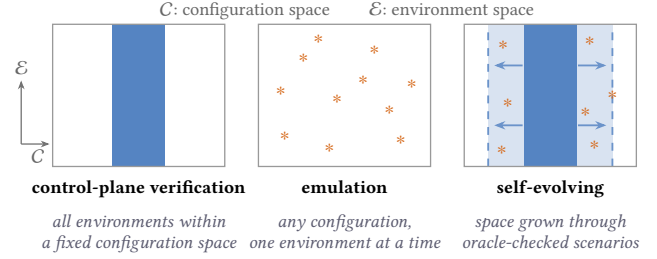

Network verification is remarkably powerful: given only a configuration, modern tools can reason about network behaviours across all possible routing inputs and link failures~\cite{minesweeper,abstract-interp}---at least, in theory.

In practice, though, a decade of academic success has not translated (yet) into (vast) industry adoption. In a recent survey of network operators, the most frequently stated deterrent is that existing tools do not support the protocols and features their networks run~\cite{krentsel-model-free}. As a result, most networks remain unverified and continue to suffer the outages that verification was built to prevent~\cite{campion}.

The reason can be traced to a single artifact: the network control-plane \emph{model}, which is today built entirely by hand.
Existing verifiers indeed support only the features their authors have manually encoded, which is only a subset of what operators actually run. This is not surprising: accounting for all the different protocols and vendor implementations is a Sisyphean task. (The maintainers of Batfish already report a decade's worth of extending vendor and protocol support and of chasing undocumented semantics in router software~\cite{batfishevolution}.) Worse, extending models is often disproportionately hard because of how protocols interact. For instance, a BGP verifier cannot ``only'' model the BGP computation across vendors and implementations; it also needs to model the IGP computation logic, again across vendors and implementations, as the BGP's route selection depends on the IGP costs~\cite{minesweeper}.

These limits have pushed recent works to make radical proposals: in~\cite{krentsel-model-free}, the authors argue that \textit{``models will always lag in their ability to faithfully represent the actual control plane''}. They propose abandoning the models altogether and instead running the configuration on an emulated network, analysing the resulting data plane directly. Such \textit{model-free} approaches overcome the limitations of handcrafted models: any feature the router image supports is covered, the image is always up-to-date, and there is no encoding to extend.

Foregoing models comes with a high cost, though: the loss of widely applicable guarantees. For a single converged data plane, we can easily reason about properties without using a model; we just run the configuration and then verify correctness over the packet space~\cite{HSA,veriflow}. But whether the network \emph{stays} correct across environments (any link failure, any external advertisement) is a question that a single data plane cannot answer. Most of the time, the environment space is huge and combinatorial, making it intractable to enumerate every possible data plane. Relying on a (symbolic) model~\cite{abstract-interp} is precisely what enables control-plane verification to reason about all environments at once.

In this paper, we argue that the problem is \textit{not} relying upon a model but that humans write it. To address this, we describe a vision in which symbolic models \textit{automatically} evolve to faithfully represent the behaviour of the network they describe (Fig.~\ref{fig:tradeoff}). In this world, models automatically cover new features, adapt to new vendors or software updates, without any expert labour. More importantly, they do so without surrendering the vast guarantees they provide.

Realizing this vision is, quite unsurprisingly, difficult. Yet, we believe the recent advances in network emulation, coupled with the advances in LLM-based code generation, make it possible. Concretely, we treat the network model like software under test: we continually compare what the model admits against what the network computes, and let an LLM coding agent repair the model wherever the two disagree.

Any procedure that changes the model automatically needs a criterion for when a change is correct~\cite{synthesis}. In networking, deriving such a specification is exceedingly difficult because no written source exists that specifies what a network actually does. Protocol implementations diverge from the RFCs~\cite{hoyan}, vendor documentation is incomplete~\cite{batfishevolution}, protocols interact in ways no standard describes~\cite{metha}, and behaviour shifts across even minor releases. The one artifact that fully determines a network's behaviour is the router software, and unlike a written specification, it can be executed: emulation can therefore answer, for any configuration in one concrete environment, exactly what the model should say.

We therefore propose to use this executable ground truth to drive a counterexample-guided synthesis loop: for each configuration and concrete environment, we check that the symbolic model is consistent with the routing state the network computed. Every disagreement becomes a counterexample, and an off-the-shelf coding agent turns counterexamples into candidate repairs (current LLMs have been shown to be effective in various adjacent domains~\cite{event-b-agent,orvalho2024counterexampleguidedprogramrepair,eywa,ma2026specsynllmbasedsynthesisrefinement}).
The specification, in short, is the network you run, not the one the standards describe: in our experiments (\S\ref{subsec:agent-models}), the agents initially followed RFC~4456's route-selection order~\cite{rfc4456}, but the ground truth followed a different order~\cite{frr-bgp-route}; counterexamples guided the agents to it.

We present early evidence that our vision works end-to-end. On a 3,000-line Rust SMT verifier, this pipeline autonomously added three mechanisms the model did not support: OSPF areas, BGP route reflection, and L3VPN over EVPN, each for a few tens of dollars and with no expert in the loop beyond inspecting the end result.

While this approach drastically reduces the cost of obtaining a \textit{possibly correct} network model, the hard problem shifts from writing verifiers to systematically testing them: what makes a corpus of tests adequate, and how far can we trust an oracle beyond the tested region? If we can confidently reason about the soundness of an automatically evolved verifier, the implications would be paramount: expanding protocol and property coverage, making verification more accessible and useful to many networks, and thereby shielding them from regularly occurring outages.

\begin{figure*}
  \input{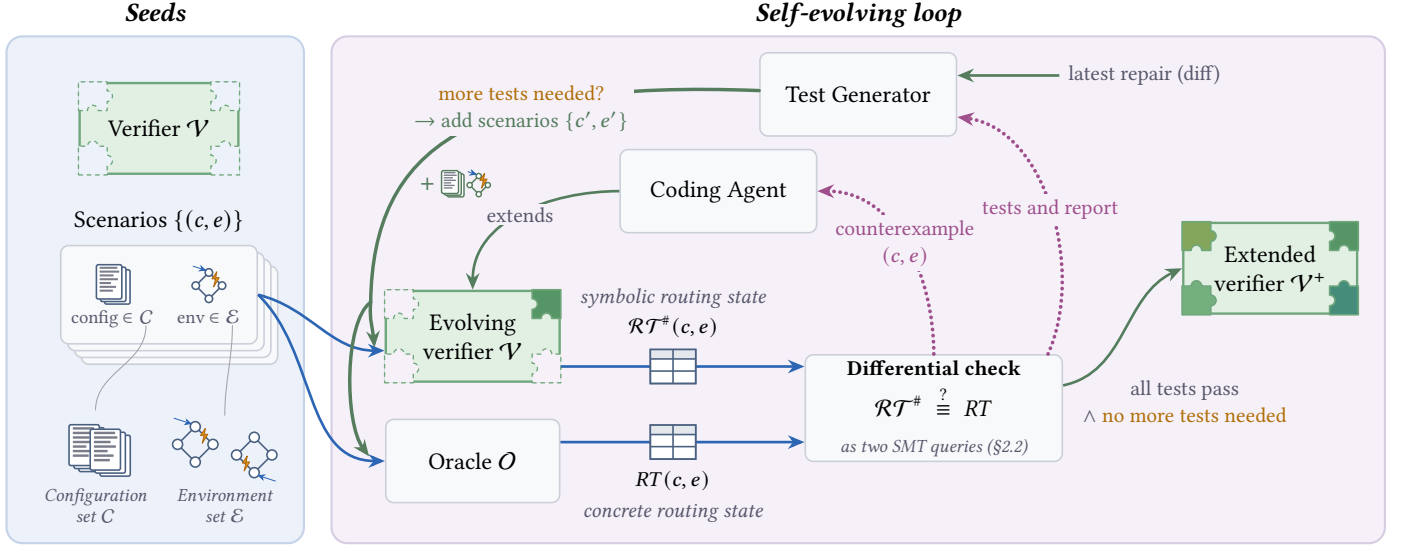}
  \caption{Every disagreement between the symbolic model and the Oracle becomes a counterexample that drives the next repair; the loop ends only when the whole corpus passes \emph{and} the Test Generator can grow it no further.}
\label{fig:loop}
\end{figure*}

\section{Self-evolving verifier}
\label{sec:system}

This section makes the vision concrete: we state the problem (\S\ref{subsec:problem}), show how to test a symbolic model against an Oracle (\S\ref{subsec:diff-test}), close the loop that repairs it (\S\ref{subsec:loop}), and state what a passing run does (and does not) certify (\S\ref{subsec:certifications}).

\subsection{Problem statement}
\label{subsec:problem}

We are given a set of configurations $\mathcal{C}$, the environments $\mathcal{E}$ they may face---link failures, external advertisements---and a verifier $\mathcal{V}$ that cannot reason correctly about every \emph{scenario} $(c,e)$: a configuration paired with one concrete environment. We also have an Oracle $\mathcal{O}$ we trust (a simulator, emulator, or the network itself), which, given a scenario, returns the routing state the network converges to. For simplicity, we consider only networks where this state is unique (otherwise, every notion of agreement below would range over all possible stable routing states~\cite{stable-paths,gao-rexford}).

Our goal is to produce an \textit{extended} verifier $\mathcal{V}^+$ that agrees with the Oracle on all scenarios. More concretely, the verifier's model must accept the routing state the Oracle produces and must accept no other. As with any tested system, if a large and varied corpus cannot surface any difference between the model and the Oracle, we trust the model across all configurations in $\mathcal{C}$ and environments in $\mathcal{E}$.

\subsection{Differentially testing a verifier}
\label{subsec:diff-test}

A network model produced by a verifier is not directly comparable to the output of a simulator or an emulator; the former is an abstraction over the network control plane across all possible environments, while the latter takes the configuration and one concrete environment as input and produces the routing state.

Still, we can query the verifier's network model to test whether the model is \textit{consistent} with the routing state calculated by the Oracle for a specific environment. That involves crafting two queries that check for two converse requirements; for a given configuration $c$ and environment $e$:

\fakepar{Soundness:} Does the generated SMT model reject a routing state that is observed through the Oracle $\mathcal{O}$?
We can effectively check soundness by crafting an SMT query that asks \textit{``Given the environment $e$, is it possible for the routes in the SMT model to be the same as the ones produced by the Oracle?''}. If we denote as $r$ each route in the routing table calculated by the Oracle $RT=\mathcal{O}(c,e)$, and $r^{\#}$ its respective symbolic route under the SMT model, then that query would be
    $$e \wedge \left( \bigwedge_{r \in RT} r = r^{\#}\right) 	\rightsquigarrow (\textit{should be } \textsc{sat})$$

Here $r=r^\#$ asserts, for a concrete route $r$ and its symbolic counterpart at the same (router, destination) slot, agreement on availability and on every route attribute (next hop, local preference, AS-path, and so on).
A \textsc{sat} result here exhibits a model assignment matching reality; \textsc{unsat} means the encoding forbids the real state.

\fakepar{Completeness:} Does the generated SMT model accept a routing state that cannot be observed through the Oracle $\mathcal{O}$? Conversely, we check completeness by querying whether the symbolic encoding can disagree with the Oracle's state.
$$e \wedge \left( \bigvee_{r \in RT} r \not= r^{\#}\right) 	\rightsquigarrow (\textit{should be } \textsc{unsat})$$
A \textsc{sat} result here would mean that the model accepts a spurious state, revealing a bug (or an omission) in the encoding.

\subsection{The evolution loop}
\label{subsec:loop}

The differential testing approach above turns a single scenario into a verdict on the current encoding, but a verdict alone does not repair the model. Two agents close the loop (Fig.~\ref{fig:loop}): a \emph{Coding Agent} that continually refines the verifier's symbolic encoding, and a \emph{Test Generator} that extends the scenario corpus.

The loop follows counterexample-guided inductive synthesis (CEGIS)~\cite{1714168}, with an LLM as the inductive synthesizer~\cite{orvalho2024counterexampleguidedprogramrepair}, and with the specification that would ordinarily bound the search replaced by the Oracle (a form of OGIS~\cite{jha2016theoryformalsynthesisinductive}). We seed the corpus with scenarios $(c,e)$ and run the differential test on each of them. A scenario in which either query fails is a counterexample: a configuration and environment in which the encoding and the Oracle disagree. We hand it to the Coding Agent along with the failing query and the Oracle's routing state; the agent edits the encoding, we recompile, and the entire corpus is re-run so that a fix for one feature cannot regress another. Iteration continues until every scenario passes.

Passing the corpus certifies the encoding only on the scenarios it contains, so it is not, on its own, a stopping condition. We therefore ask the Test Generator whether any untested region of the configuration space remains. If it can produce a new scenario that the encoding and Oracle have not yet been checked against, that scenario enters the corpus, and the loop resumes; only when all tests pass, and no further scenario is needed, do we emit the extended verifier $\mathcal{V}^+$.

\subsection{What a passing run means}
\label{subsec:certifications}

A successful test run certifies that, for every scenario in the corpus, the verifier encoding admits the Oracle's routing state, and no other. Any claim beyond the corpus is inductive, so naturally, there can still be uncaught errors; untested regions of the configuration and environment space may still hide modelling bugs. Furthermore, the symbolic model remains an abstraction of actual network behaviour, meaning that properties more granular than this abstraction cannot be tested against an emulator.

While we recognize that this is weaker than a proof, hand-written encodings do not come with proofs either; they are validated with finite test suites, expert review, and hand-maintained comparisons against ground truth (e.g., Batfish against emulated networks~\cite{batfishevolution}, Alibaba's Hoyan continuously against its live WAN~\cite{hoyan2}), and prominent verifiers ship fidelity bugs regardless \cite{metha, krentsel-model-free}. So, irrespective of the paradigm that creates the verifier, there is a need to systematize the testing of such systems, and differential testing~\cite{McKeeman1998DifferentialTF} against---now cheaper and more accessible---emulators is a step in that direction.

\section{Early results}
We implemented and ran the loop of \S\ref{sec:system} on an SMT-based control-plane verifier ($\sim$3000 lines of Rust using Z3~\cite{z3}) that models BGP over a flat shortest-path IGP. We gave it three extension tasks for mechanisms it did not model: \textbf{(i)~OSPF areas}, \textbf{(ii)~BGP route reflection} (RFC\,4456), including cluster lists, originator IDs, and their effect on best-path selection; and \textbf{(iii)~L3VPN over EVPN}, a feature family the verifier had no notion of. The oracle is hot-swappable, and we used two oracles of different fidelity: \textit{Batfish}~\cite{batfish} for (i) and (ii), and emulated \textit{FRRouting} instances (via containerlab) for (iii), where L3VPN support is more complete. A general \emph{Coding Agent} (Claude Opus 4.8 or Sonnet~5) is tasked with implementing the update to the existing symbolic verifier; the \textit{Test Generator} is a separate agent (Opus 4.8 throughout) that reads the implementer's code and existing test cases each round and proposes new scenarios.

Table~\ref{tab:runs} summarizes the runs. For tens of dollars and $\sim$100--250 agent steps per feature, every run converged, via oracle counterexamples and internal protocol knowledge, to a verifier that fully agreed with its oracle on a test corpus adversarially grown by the Test Generator. Not every model we tried sufficed: runs with Haiku~4.5 as the Coding Agent failed to converge on any of the three tasks.

\begin{table}[t]
  \scriptsize
  \setlength{\tabcolsep}{2.3pt}
  \begin{tabular}{@{}lcccccc@{}}
    \toprule
     & \multicolumn{2}{c}{\textbf{OSPF areas}} & \multicolumn{2}{c}{\textbf{Route refl.}} & \multicolumn{2}{c}{\textbf{L3VPN}} \\
    \cmidrule(lr){2-3}\cmidrule(lr){4-5}\cmidrule(l){6-7}
    Coding Agent model   & Opus & Sonnet & Opus & Sonnet & Opus & Sonnet \\
    \midrule
    Tests (seed$\to$final) & 316$\to$544 & 316$\to$648 & 484$\to$600 & 484$\to$600 & 508$\to$596 & 508$\to$582 \\
    Rounds (failing) & 7 (3) & 9 (1) & 5 (1) & 4 (0) & 13 (3) & 11 (2) \\
    Agent steps   & 95 & 91 & 200 & 173 & 218 & 254\\
    Total cost (\$)  & 41.8 & 35.8 & 30.1 & 18.7 & 37.2 & 25.0 \\
    Lines added      & 288 & 275 & 85 & 152 & 1162 & 1060 \\
    Final passing tests        & 544/544 & 648/648 & 600/600 & 600/600 & 596/596 & 582/582 \\
    \bottomrule
  \end{tabular}
  \caption{Every run converged to full agreement with its oracle on an implementation-aware testing suite.}
  \label{tab:runs}
\end{table}

\subsection{What the agents modelled}
\label{subsec:agent-models}

Figure~\ref{fig:encodings} distills each converged encoding into the rules the agent effectively added, as faithful abridgements of the agents' Rust/Z3 code.

\fakepar{OSPF areas (Fig.~\ref{fig:encodings}i).} Both agents independently converged on the same formulation. They defined symbolic per-area distances and composed them into inter-area distances dependent on link availability. Correctly modelling shortest-path calculation under the partitioning of Area 0 was a subtlety that required both agents to refine their code through counterexamples.

\fakepar{Route reflection (Fig.~\ref{fig:encodings}ii).} Both agents
converged on FRR's decision process, even on a detail that deviates from the route reflection RFC: in \textit{Batfish}, FRR compares cluster-list length right after the IGP-metric step, ahead of the originator-ID comparison~\cite{batfish-bgprib}, whereas RFC~4456~\S9 prescribes the opposite order. Originally, the agents implemented an SMT model from their knowledge of the protocol, so the route decision process followed the RFC. They adapted to the FRR logic only through counterexamples authored by the Test Generator.

\fakepar{L3VPN over EVPN (Fig.~\ref{fig:encodings}iii).} Both agents converged on the encoding shown, including all eight terms of the decision process $\prec$. There were, however, subtleties in the network that the RFCs do not specify: a prefix advertised from two sites has a different best route at its origin than in the rest of the network, because a router privately prefers its own routes through an attribute that never leaves the router. The router software had to reveal such behaviour~\cite{frr-bgp-route}.

\begin{figure}[t]
  \centering
  \scriptsize

  \newcommand{\enckw}[1]{\textcolor{titlegray}{\scriptsize\textsc{#1}}}
  \newcommand{\encsub}[1]{\textcolor{titlegray}{\scriptsize #1}}
  \newcommand{\enccmt}[1]{\textcolor{inkblue}{\scriptsize\itshape $\triangleright$\,#1}}
  \newcommand{\encnote}[1]{\textcolor{duskpurple}{\scriptsize $\hookrightarrow$\,#1}}
  \setlength{\tabcolsep}{0pt}
  \renewcommand{\arraystretch}{1.15}
  \begin{tabular}{@{}p{0.85cm}@{\hspace{4pt}}p{\dimexpr\columnwidth-0.85cm-4pt\relax}@{}}
    \toprule
    \multicolumn{2}{@{}p{\columnwidth}@{}}{\textbf{(i) OSPF areas}}\\
    \addlinespace[3pt]
    \enckw{define} & per-area distances $d_a(u,v)$: symbolic shortest paths over the area-$a$ links that are up in the environment \\
    \addlinespace[2pt]
    \enckw{define} & $D(s,t) \;=\; \displaystyle\min_{\substack{x \,\in\, \mathrm{ABR}(a_s)\\ y \,\in\, \mathrm{ABR}(a_t)}}\; d_{a_s}(s,x) + d_{0}(x,y) + d_{a_t}(y,t)$ \newline
    \encnote{Despite leading to the exact same result, Opus considered the min over a much vaster space (\S\ref{subsec:performance})} \\
    \addlinespace[2pt]
    \enckw{assert} & each path segment is reachable within its area \\
    \addlinespace[1pt]
    \enckw{assert} & an intra-area route, when one exists, beats every inter-area route, regardless of cost \\
    \addlinespace[3pt]
    \midrule
    \multicolumn{2}{@{}p{\columnwidth}@{}}{\textbf{(ii) Route reflection}\;}\\
    \addlinespace[3pt]
    \enckw{define} & per-route attributes: cluster list $\mathit{CL}$, originator ID $\mathit{OID}$ \\
    \addlinespace[2pt]
    \enckw{define} & $\mathit{CL} \leftarrow \mathit{cid}(r) \,{::}\, \mathit{CL}$ \hfill \enccmt{reflector $r$ prepends its cluster ID} \\
    \addlinespace[1pt]
    \enckw{define} & $\mathit{OID} \leftarrow \textit{sender}$ if unset \hfill  \\
    \addlinespace[1pt]
    \enckw{assert} & router $t$ admits a route iff\; $\mathit{OID} \neq t \,\wedge\, cid(t) \notin \mathit{CL}$ \hfill \enccmt{loop rejection} \\
    \addlinespace[2pt]
    \enckw{define} & lexicographic minimization along \hfill \enccmt{best-path selection} \newline
    $\cdots \,\succ\, \text{IGP cost}  \,\succ\, | \mathit{CL} | \,\succ\, \mathit{OID} \,\succ\, \text{peer ID}$ \newline
    \\
    \addlinespace[3pt]
    \midrule
    \multicolumn{2}{@{}p{\columnwidth}@{}}{\textbf{(iii) L3VPN over EVPN}}\\
    \addlinespace[3pt]
    \enckw{given} & routers $R$, weighted core links $L$, EVPN sessions $E$, VRF attachments, originated prefixes $O$ \\
    \addlinespace[2pt]
    \enckw{define} & $d(u,v)$: shortest paths after deleting failed nodes and links \\
    \addlinespace[2pt]
    \enckw{define} & $\mathit{act}(u,v) \,\Leftrightarrow\, (u,v)\in E \,\wedge\, \text{both up} \,\wedge\, d(u,v)<\infty$ \newline
    \mbox{}\hfill \enccmt{EVPN session liveness} \\
    \addlinespace[1pt]
    \enckw{define} & $o \rightsquigarrow t$: RFC~4456 reflection reachability over $\mathit{act}$ \hfill \enccmt{overlay RR} \\
    \addlinespace[1pt]
    \enckw{define} & $\mathit{rt}(v) =$ ASN:VNI if auto,\quad $\mathit{rd}(v) =$ routerID:vrfID if auto \newline
    \mbox{}\hfill \enccmt{route target / distinguisher} \\
    \addlinespace[1pt]
    \enckw{define} & a type-5 route is $\langle \mathit{rd}(o),\, p,\, \mathit{RT}(o),\, \mathcal{A}(o)\rangle$ \hfill \enccmt{RD keys identity} \\
    \addlinespace[1pt]
    \enckw{define} & $\prec \;=\; \big(\neg\mathit{weight},\, \mathrm{LP},\, |\mathrm{ASpath}|,\, \mathrm{origin},\, \mathrm{MED},\, \mathrm{eBGP}{\prec}\mathrm{iBGP},$ \newline
    \hspace*{1.6em}$\text{IGP cost},\, \text{router ID}\big)$ \hfill \enccmt{full decision process} \\
    \addlinespace[1pt]
    \enckw{define} & $\mathit{ebest}(r,\mathit{rd},p) = \min_{\prec}\,\{\, o \,:\, \mathit{rd}(o)=\mathit{rd} \,\wedge\, d(r,o)<\infty \,\wedge\, o \rightsquigarrow r \,\}$ \newline
    \mbox{}\hfill \enccmt{EVPN RIB: selection per (RD, prefix), \emph{not} across RDs} \\
    \addlinespace[1pt]
    \enckw{define} & $\mathit{cand}(r,v,p) = \{\, \mathit{ebest}(r,\mathit{rd},p) \,:\, \mathit{rt}^{\mathrm{imp}}(r,v) \in \mathit{RT}(o)$ \newline
    \hspace*{5.4em}$\wedge\;\, \mathit{rd}(o) \neq \mathit{rd}(v) \,\}$ \hfill \enccmt{leak: import filter + self-RD rejection} \\
    \addlinespace[1pt]
    \enckw{define} & $\mathit{best}(r,v,p) = \min_{\prec}\, \mathit{cand}(r,v,p)$ \newline
    \mbox{}\hfill \enccmt{tenant v4 RIB: cross-RD competition resolved here} \\
    \bottomrule
  \end{tabular}
  \caption{The two agents converged to semantically equivalent encodings.}
  \label{fig:encodings}
\end{figure}

\subsection{Tests must co-evolve with the model}

\begin{figure}
  \centering

\definecolor{cpass}{HTML}{2B65B3}
\definecolor{cfail}{HTML}{C43C3C}
\definecolor{cconv}{HTML}{0A7A0A}
\pgfplotsset{
  cegispanel/.style={
    scale only axis=true, width=3.05cm, height=2.05cm,
    axis lines=left, axis line style={-, line width=0.5pt, black!70},
    tick align=outside, tick style={black!70}, major tick length=1.8pt,
    xmin=-0.6, xmax=9.6, xtick={0,3,6,9},
    x tick label style={font=\scriptsize},
    y tick label style={font=\scriptsize},
    label style={font=\scriptsize},
    ymajorgrids, grid style={black!12, line width=0.3pt},
    clip=false,
    every axis plot/.append style={line width=0.9pt},
  },
  toprow/.style={ymin=150, ymax=740, ytick={200,400,600}},
  botrow/.style={ymin=380, ymax=660, ytick={450,525,600}},
  pass/.style={cpass, solid, mark=*, mark size=1.15pt, mark options={fill=cpass, draw=cpass}},
  conv/.style={cconv, only marks, mark=*, mark size=1.7pt},
  xmark/.style={cfail, only marks, mark=x, mark size=2.2pt,
                mark options={line width=0.8pt, draw=cfail}},
}
\tikzset{
  faillab/.style={cfail, font=\scriptsize, inner sep=1.5pt},
  convlab/.style={cconv, font=\scriptsize\bfseries, inner sep=1.5pt, yshift=1.5pt},
  failtick/.style={cfail, line width=0.8pt},
}

\begin{tikzpicture}[font=\scriptsize, baseline=-0.5ex]
  \draw[cpass, line width=0.9pt] (0,0)--(0.4,0);
  \fill[cpass] (0.2,0) circle (1.1pt);
  \node[anchor=west, inner sep=2pt] at (0.42,0) {passing};
  \draw[cfail, line width=0.8pt] (1.75,-0.09)--(1.75,0.05);
  \node[cfail, font=\scriptsize, inner sep=0pt] at (1.75,0.09) {$\times$};
  \node[anchor=west, inner sep=2pt] at (1.87,0) {failing (count)};
  \fill[cconv] (3.95,0) circle (1.6pt);
  \node[anchor=west, inner sep=2pt] at (4.07,0) {converged};
\end{tikzpicture}

\medskip

\begin{subfigure}{0.48\linewidth}\centering
\begin{tikzpicture}
\begin{axis}[cegispanel, toprow,
  ylabel={tests passed}, ylabel style={yshift=-4pt}]
\addplot[pass] coordinates
  {(0,216)(1,216)(2,360)(3,400)(4,444)(5,480)(6,544)(7,544)};
\addplot[conv] coordinates {(7,544)};
\draw[failtick] (axis cs:0,216) -- (axis cs:0,316);
\draw[failtick] (axis cs:1,216) -- (axis cs:1,316);
\node[faillab, anchor=south] at (axis cs:0.5,332) {100};
\node[faillab, anchor=north] at (axis cs:4,428) {4};
\node[faillab, anchor=north] at (axis cs:5,464) {8};
\node[convlab, anchor=south] at (axis cs:7,544) {544/544};
\addplot[xmark] coordinates {(0,316)(1,316)(4,452)(5,492)};
\end{axis}
\end{tikzpicture}
\caption{OSPF areas $\cdot$ Opus}\label{fig:cegis-ospf-opus}
\end{subfigure}\hfill
\begin{subfigure}{0.48\linewidth}\centering
\begin{tikzpicture}
\begin{axis}[cegispanel, toprow, yticklabels={}]
\addplot[pass] coordinates
  {(0,216)(1,348)(2,416)(3,456)(4,480)(5,528)(6,580)(7,624)(8,648)(9,648)};
\addplot[conv] coordinates {(9,648)};
\draw[failtick] (axis cs:0,216) -- (axis cs:0,316);
\node[faillab, anchor=south] at (axis cs:0.1,332) {100};
\draw[failtick] (axis cs:1,348) -- (axis cs:1,368);
\node[faillab, anchor=south] at (axis cs:0.72,392) {20};
\node[convlab, anchor=south] at (axis cs:9,648) {648/648};
\addplot[xmark] coordinates {(0,316)(1,372)};
\end{axis}
\end{tikzpicture}
\caption{OSPF areas $\cdot$ Sonnet}\label{fig:cegis-ospf-sonnet}
\end{subfigure}

\medskip

\begin{subfigure}{0.48\linewidth}\centering
\begin{tikzpicture}
\begin{axis}[cegispanel, botrow,
  ylabel={tests passed}, ylabel style={yshift=-4pt}, xlabel={round}]
\addplot[pass] coordinates
  {(0,433)(1,508)(2,550)(3,576)(4,600)(5,600)};
\addplot[conv] coordinates {(5,600)};
\draw[failtick] (axis cs:0,433) -- (axis cs:0,484);
\node[faillab, anchor=north] at (axis cs:0,418) {51};
\draw[failtick] (axis cs:2,550) -- (axis cs:2,552);
\node[faillab, anchor=south] at (axis cs:2,560) {2};
\node[convlab, anchor=south] at (axis cs:5,600) {600/600};
\addplot[xmark] coordinates {(0,484)(2,552)};
\end{axis}
\end{tikzpicture}
\caption{Route reflection $\cdot$ Opus}\label{fig:cegis-rr-opus}
\end{subfigure}\hfill
\begin{subfigure}{0.48\linewidth}\centering
\begin{tikzpicture}
\begin{axis}[cegispanel, botrow, yticklabels={}, xlabel={round}]
\addplot[pass] coordinates
  {(0,433)(1,540)(2,560)(3,600)(4,600)};
\addplot[conv] coordinates {(4,600)};
\draw[failtick] (axis cs:0,433) -- (axis cs:0,484);
\node[faillab, anchor=north] at (axis cs:0,418) {51};
\node[convlab, anchor=south] at (axis cs:4,600) {600/600};
\addplot[xmark] coordinates {(0,484)};
\end{axis}
\end{tikzpicture}
\caption{Route reflection $\cdot$ Sonnet}\label{fig:cegis-rr-sonnet}
\end{subfigure}
  \caption{An evolved verifier that passes all tests in a single round might still fail on different inputs.}
  \label{fig:cegis-progress}
\end{figure}
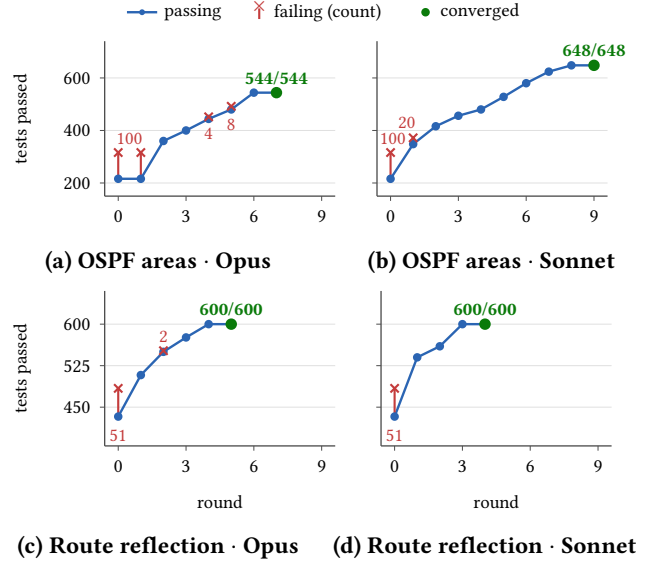

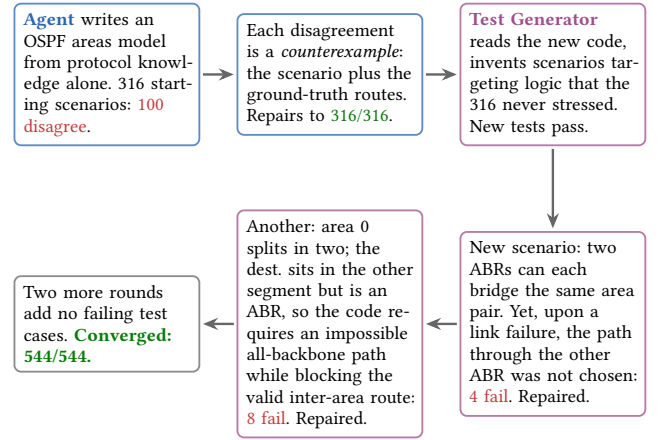
\begin{figure}[t]
  \centering

\begin{tikzpicture}[
  font=\scriptsize,
  evbox/.style={rounded corners=2.5pt, draw, line width=0.7pt, fill=white,
                text width=2.2cm, align=left, inner sep=3.5pt},
  pbox/.style={evbox, draw=inkblue!75},
  cbox/.style={evbox, draw=duskpurple!75},
  gbox/.style={evbox, draw=black!45},
  stepline/.style={->, line width=0.9pt, black!60, >={Stealth[length=2mm,width=1.6mm]}, shorten >=1pt, shorten <=1pt},
]
\definecolor{okgreen}{HTML}{0A7A0A}
\definecolor{badred}{HTML}{C43C3C}
\node[pbox] (b1) at (0,0)
  {\textcolor{inkblue}{\textbf{Agent}} writes an OSPF areas model from protocol knowledge alone. 316 starting scenarios: \textcolor{badred}{100 disagree}.};
\node[pbox] (b2) at (2.95,0)
  {Each disagreement is a \emph{counterexample}: the scenario plus the ground-truth routes. Repairs to \textcolor{okgreen}{316/316}.};
\node[cbox] (b3) at (5.9,0)
  {\textcolor{duskpurple}{\textbf{Test Generator}} reads the new code, invents scenarios targeting logic that the 316 never stressed. New tests pass.};
\node[cbox] (b4) at (5.9,-3.3)
  {New scenario: two ABRs can each bridge the same area pair. Yet, upon a link failure, the path through the other ABR was not chosen: \textcolor{badred}{4 fail}. Repaired.};
\node[cbox] (b5) at (2.95,-3.3)
  {Another: area~0 splits in two; the dest. sits in the other segment but is an ABR, so the code requires an impossible all-backbone path while blocking the valid inter-area route: \textcolor{badred}{8 fail}. Repaired.};
\node[gbox] (b6) at (0,-3.3)
  {Two more rounds add no failing test cases. \textcolor{okgreen}{\textbf{Converged: 544/544.}}};
\draw[stepline] (b1) -- (b2);
\draw[stepline] (b2) -- (b3);
\draw[stepline] (b3) -- (b4);
\draw[stepline] (b4) -- (b5);
\draw[stepline] (b5) -- (b6);
\end{tikzpicture}
  \caption{The Test Generator proposes adversarial scenarios until it cannot invent further ones.}
  \label{fig:looptrace}
\end{figure}

Figure~\ref{fig:cegis-progress} traces how a run's corpus and
passing checks move together: in each round, the Test Generator
tries to grow the corpus with adversarial examples, and each red
cross is a round that ended with counterexamples. The runs share a
common structure. The verifier's initial state fails a large fraction of the seed corpus---unsurprisingly so, since a lot of the seed scenarios exercise the very feature being added (e.g., 100 of 316 for OSPF areas)---and the agent
repairs against those counterexamples until the seed passes. Had the
loop stopped there, the model would \emph{look} finished. Yet in
every run, the Test Generator, reading the converged code, invented
scenarios the seed never stressed, and some promptly failed.
Figure~\ref{fig:looptrace} follows one such trajectory in detail.

Passing the current corpus is thus a fixed point of \emph{repair}, not of correctness: each repaired model has behaviour that only fresh, code-aware tests can probe. Our Test Generator finds such tests, but greedily and without guarantees: nothing tells us how far ``the Test Generator ran out of ideas'' is from ``no counterexample exists.'' Closing that gap, e.g. by generating adversarial scenarios systematically, with some notion of coverage over a targeted scenario space, is, we argue, a core open problem for self-evolving verifiers (\S\ref{sec:research-agenda}).

\subsection{Performance is important too!}
\label{subsec:performance}
Consistency with the oracle is the main signal that drives the loop
of \S\ref{sec:system}. But a verifier has (at least) two different
requirements: \emph{correctness}, a hard constraint that admits no
tradeoff, and \emph{performance}, a soft objective the model can
satisfy to wildly varying degrees. The oracle enforces only the
first on the test corpus; computational tractability is invisible to it.

We benchmarked each pair of converged network models on the same
query---proving reachability under all single-link failures---over
topologies at growing sizes. Each pair yields the same verdict, yet
their costs diverge by orders of magnitude. For OSPF areas, Opus's
encoding considers vastly more candidate inter-area paths than are
actually possible, while Sonnet derives each area's border routers
from the configuration and admits only paths through them. The
redundant computations compound with scale: at just 18 routers, Opus's encoding
cannot answer the query within 20 minutes, while Sonnet's takes
0.16\,s. For L3VPN at 17 routers, Sonnet's
model takes 143\,s, whereas Opus's takes 0.9\,s (all measurements are means over 10 runs).

To test whether the loop can also pursue the performance objective,
we resumed the converged Opus OSPF run with a single generic
instruction (the encoding must also scale) and a cheap size signal
(SMT DAG nodes), with no hint at the underlying anti-pattern. In two
rounds the agent independently reinvented the ABR restriction:
correctness held at every step (544/544 against the oracle), and the
once-intractable query now completes in 0.4\,s. A tangible gap
remains, however: Sonnet's encoding is still over $2\times$
faster on the same proof, so even within verifiers with similar network models and the same end results, performance may vary significantly.

\section{Research Agenda}
\label{sec:research-agenda}

\fakepar{Systematic testing of verifiers.} The system in \S\ref{sec:system} drastically cuts the cost of writing a verifier, leaving only the task of rigorously testing one. The need is not ours alone: the operators of Alibaba's production verifier explicitly call for an automated framework that tests a verifier against vendor implementations and diagnoses the causes of divergence~\cite{hoyan2}. While our Test Generator does expose failures by probing logic that the previous test suite never stressed, it cannot guarantee that every corner case is covered.

Tests must cover two spaces: configurations and network environments. Regarding configurations, we need ones that surface discrepancies between the network model and ground-truth behaviour, and such scenarios are sparse: validating the order of a decision process at step $k$ requires routes with equal attributes on all previous steps, as in the route-reflection tie-break of \S\ref{subsec:agent-models}. Our Test Generator greedily finds such scenarios, with no notion of exhaustion. Recent work~\cite{aichilles} shows adversarial inputs uncover correctness and performance bugs in AI-evolved programs (e.g.,~\cite{alphaevolve,FunSearch2023}) that their original evaluation suites miss.
Similarly, an evolved verifier should be tested on adversarial scenarios.

Regarding environments, the verifier's guarantees must hold over all of $\mathcal{E}$, but each differential test checks a single point of it (the same limitation that \textit{model-free} verification faces~\cite{krentsel-model-free}). Coverage metrics for network configuration tests exist~\cite{netcov}, but they do not extend to network models that reason symbolically over environment spaces. Formal verification suggests what such a metric could look like: in model checking, a part of the model is considered covered only if mutating it reverses the verdict of some check ~\cite{coverage-metrics,coverage-estimation}. Transferred here, if mutating a piece of the model changes no scenario's outcome, then the tests never exercised that piece of logic. 

A fundamental research question remains: Can we prove a verifier to reason correctly across \textit{all} environments and supported configurations, given cheap access to the ground truth but only for one scenario at a time?

\fakepar{Optimizing for performance.} The oracle enforces correctness on the corpus, yet computational tractability is invisible to it, and \S\ref{subsec:performance} showed that verifiers that pass the same tests can differ in efficiency by orders of magnitude. A recurring cause is the boundary between what can be precomputed from the configuration (e.g., an area's border routers) and what must stay symbolic (distances under failures)~\cite{bonsai}. That boundary moves with the query: a reachability check on a fixed topology has no need for symbolic distances. Taken to its limit, concretizing everything is simply enumeration (one emulation per environment) which, for spaces like all single-link failures, may well beat a symbolic query, and certainly beats an unfortunate SMT encoding (\S\ref{subsec:performance}). If deriving a model is cheap, a verifier need not fix this boundary once; it could adapt specific calculations to use full symbolic reasoning or plain enumeration.

\fakepar{Reasoning about more properties.} The loop evolves toward whatever its oracle tests against. In this case, converged routing state (\S\ref{subsec:diff-test}). However, operators also care about transient states~\cite{schneider2023taming}, and performance properties still largely remain out of reach: general formal models for network performance scale poorly~\cite{formal-perf}, and verifying specific properties, such as worst-case link loads, requires modelling and optimizing entire systems by hand~\cite{velo}. Emulation already observes transient states, link loads, and convergence times, so the question remains: how can we use these signals to effectively steer the creation of a faithful and tractable network model? 

\fakepar{Universal or ad-hoc verifiers?} If a faithful network model costs days and tens of dollars instead of months of expert labour, what should the community maintain: one ``universal'' verifier, or a small verifier per network, covering exactly its configuration and re-evolving as it changes? Our results favour the latter, since each verifier automatically adapts to vendor-specific behaviour, whereas general configuration analysis tools like \textit{Batfish} must encode all possible vendor-dependent behaviours and select the applicable one per configuration. What the universal verifier offers instead is more comprehensive validation: every scenario tests the same code, whereas a per-network verifier is tested only in its own domain. Yet a hard scenario found by one network's loop is an oracle-checked, reusable test for any verifier that covers the same features. A community corpus of such scenarios would give ad-hoc verifiers the shared validation that universal ones enjoy today.

\bibliographystyle{ACM-Reference-Format} 
\bibliography{hotnets25-template}

\end{document}